\documentclass{PoS}
\newcommand\srm{\scriptscriptstyle\rm}
\usepackage{url}
\def\MgII{Mg\,${\srm II}$}
\def\civ{C\,${\srm IV}$}
\def\ciii{C\,${\srm III}$]}
\def\CIII{C\,${\srm III}$]}
\def\FeII{Fe\,${\srm II}$}

\title{Probing the inner structure of distant AGNs with gravitational lensing}

\ShortTitle{AGN microlensing}

\author{\speaker{D.\,Sluse}\thanks{Supported by the Deutsche Forschungsgemeinschaft, grant SL 172/1-1}\\
        Argelander-Institut f\"ur Astronomie, Auf dem H\"ugel 71, 53121 Bonn, Germany\\
        E-mail: \email{dsluse@astro.uni-bonn.de}}

\author{D.\,Hutsem\'ekers \\ FRS-FNRS;  Institut d'Astrophysique et
  de G\'eophysique, Universit\'e de Li\`ege, All\'ee du 6 Ao\^ut 17,
  B5c, 4000 Li\`ege, Belgium }

\author{F.\,Courbin \\  Laboratoire d'Astrophysique, Ecole
  Polytechnique F\'ed\'erale de Lausanne (EPFL), Observatoire de
  Sauverny, 1290 Versoix, Switzerland }

\author{G.\,Meylan \\  Laboratoire d'Astrophysique, Ecole
  Polytechnique F\'ed\'erale de Lausanne (EPFL), Observatoire de
  Sauverny, 1290 Versoix, Switzerland }

\author{J.\,Wambsganss \\ Astronomisches Rechen-Institut am Zentrum f\"ur Astronomie
  der Universit\"at Heidelberg M\"onchhofstrasse 12-14, 69120
  Heidelberg, Germany}

\abstract{Microlensing is a powerful technique which can be used to study the continuum and the broad line emitting regions in distant AGNs. After a brief description of the methods and required data, we present recent applications of this technique. We show that microlensing allows one to measure the temperature profile of the accretion disc, estimate the size and study the geometry of the region emitting the broad emission lines.}

\FullConference{Seyfert2012: Nuclei of Seyfert galaxies and QSOs - Central engine \& conditions of star formation,\\
		November 6-8, 2012\\
		Max-Planck-Insitut f\"ur Radioastronomie (MPIfR), Bonn,  			Germany}

\begin{document}

\section{Introduction}

Despite numerous progresses on our understanding of AGNs since their discovery 50 years ago, we still do not know in detail the structure of their inner region. The global picture remains fuzzy{\footnote{Although a panchromatic view of the structure is needed to understand completely AGNs, we focus hereafter on the emission in the near UV-optical range.}}. It is generally accepted that there is a central black hole which accretes matter in form of a disc. Surrounding this region, gas clouds give rise to the broad emission lines. When we try to give more details about the accretion $-$e.g. does the accretion work as imagined by Shakura and Sunyaev 40 years ago ?$-$, about the size of these regions or about the geometry and kinematics of the broad line region (BLR), we end up with elusive answers. What hampers our understanding of AGNs is the difficulty to interpret the observational signals. On one side, there is a great diversity of properties that we hardly organize. On the other side, we have only indirect observations. We would dream to image the central engine of AGNs with a sub-parsec spatial resolution, but the prospects are not encouraging. At redshift $z=3$, adaptive optics techniques are able to reveal structures on scales of 10\,mas (about 80\,pc). Interferometry in the Near-Infrared (VLTI) allows us to probe local AGNs on parsec scale (down to $10^{-1}$\,pc). An improvement by a factor 10 of the spatial resolution may be expected from upcoming observing facilities. However, this would still leave the BLR and the accretion disc mostly unresolved. Reverberation mapping is one of the roads which allows us to make progress on this issue (e.g.\cite{Peterson1993}). We discuss hereafter how cosmic microlensing enables us to scan distant quasars on micro-arcsec scales. Although this is not an imaging technique, it nicely complements existing methods such as reverberation mapping, and could in principle even be combined with that one~\cite{Garsden2011}.


Cosmic microlensing (or quasar microlensing) takes place in AGNs located ``behind'' an intervening galaxy \cite{Wambsganss2006}. Strong gravitational lensing produces multiple images of the AGN separated on the sky by a few arcsec. Microlensing is a small scale strong lensing effect produced by the stars in the lensing galaxy. Because different parts of the galaxy appear in projection towards the macro-images, each lensed image is independently affected by microlensing. Microlensing splits a macro-image in multiple images separated by micro-arcsecs. These micro-images are unresolved and only an amplification of the flux is observed. Microlensing events are frequent and generally produced by several stars on the line of sight towards a macro-image (see Sect.~\ref{sec:method}). Because of the relative motion of the stars and of the source, the microlensing phenomenon varies on scales of months to years. In addition, it magnifies differently the emitting regions in the AGN as a function of their size. Emission arising from regions with projected sizes smaller than the average Einstein radius -i.e. the lensing cross section- of the microlenses are much more affected by microlensing than regions with sizes similar or larger than the Einstein radius. Fortunately, the Einstein radius matches approximately the size of the broad line region, allowing us to study the latter as well as the more compact continuum emission. Observationally, the selective amplification from microlensing may be detected by comparing the spectra of the multiple images of a lensed AGN. A flat spectral ratio between two images indicates that those images are identical up to a multiplicative factor (the strong-lensing magnification). Instead, if microlensing affects the continuum of one image, an imprint of the broad and narrow emission lines will be visible in the spectrum ratio, because of the larger size of thes corresponding emitting regions compared to the region emitting the continuum. 



\section{Methodology}
\label{sec:method}

\begin{figure} 
\begin{tabular}{ll}
  \includegraphics[scale=0.40]{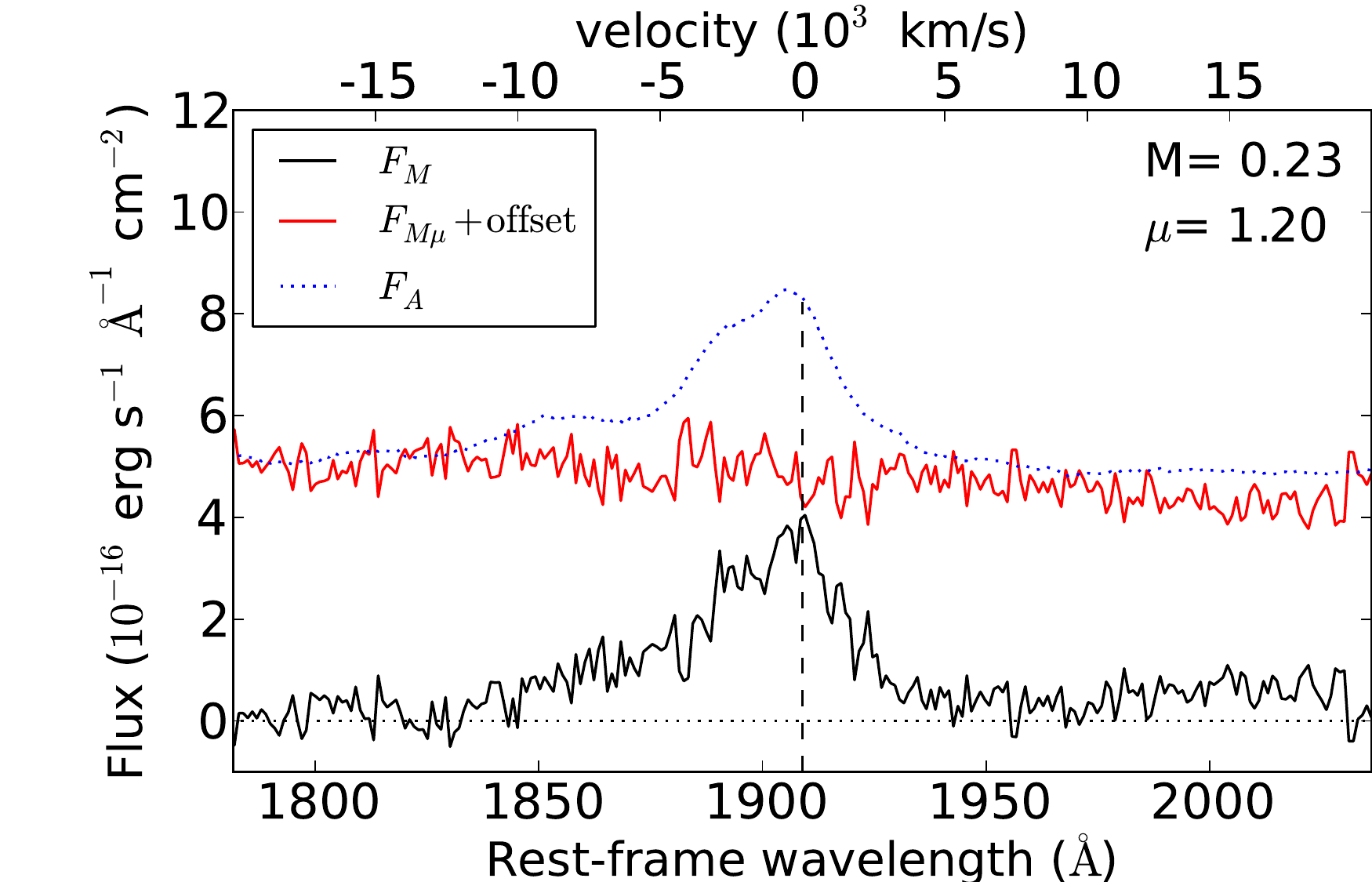} & \includegraphics[scale=0.40]{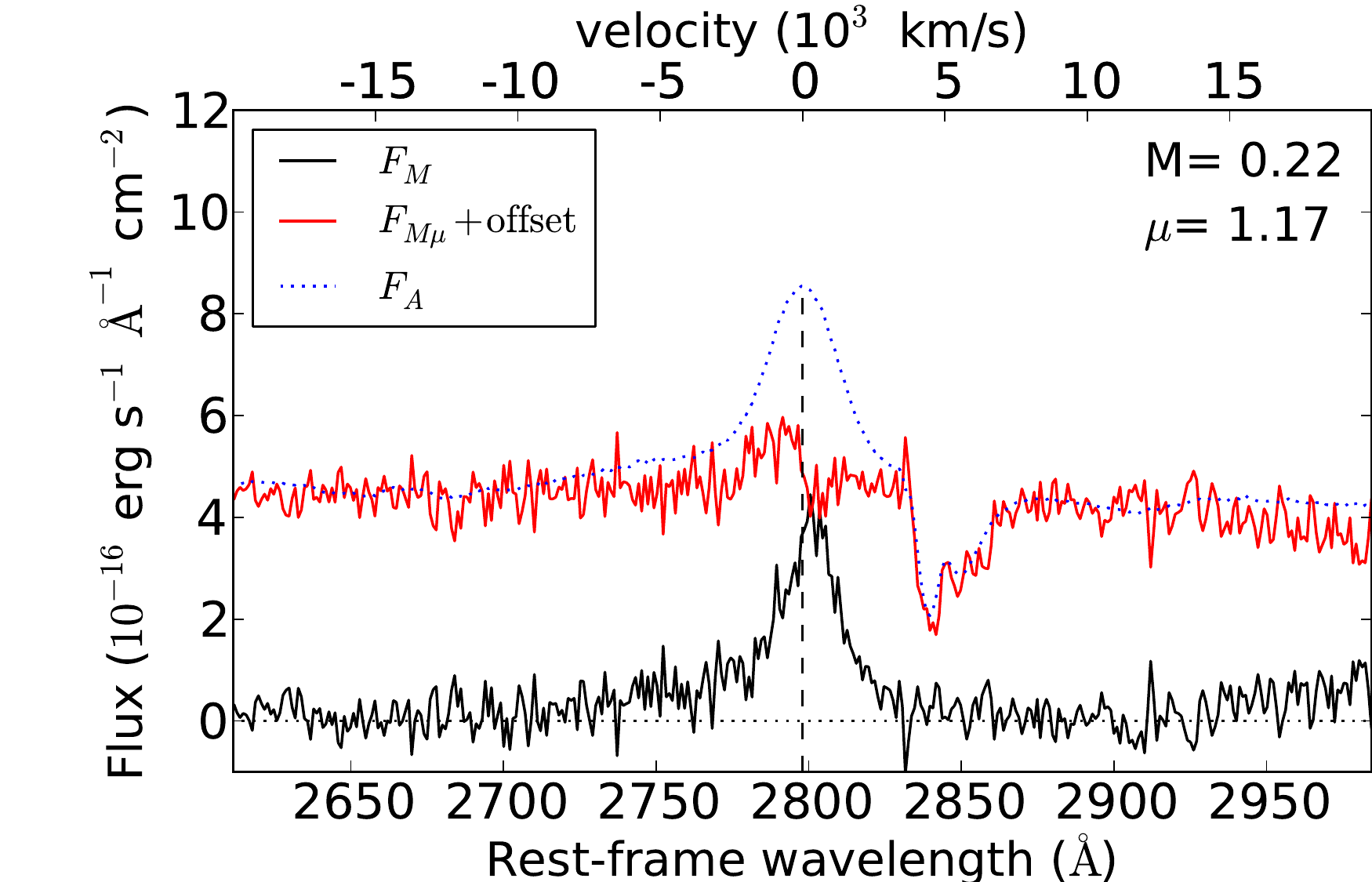} \\     
\end{tabular}
\caption{Example of Macro-micro decomposition (MmD) applied to the \ciii~line (left), and \MgII~line (right) of the lensed quasar HE0047-1756. We show the part of the line profile which is only macro-lensed, $F_M$ (black), and the part of the line profile which is both macro- and micro-lensed, $F_{M\mu}$ (red). The dotted blue line is the observed spectrum, and is equal to $F_M+F_{M\mu}$. The vertical dotted line denotes the rest wavelength of the emission line. Microlensing of the blue wing of \MgII~is clearly visible (as a bump in $F_{M\mu}$). There is also a hint of for a similar effect in \ciii.}
\label{fig:MmD}
\end{figure}

Calculating a spectral ratio is the most basic test one can perform to unveil microlensing in lensed AGNs' spectra. Another technique, named MmD for Macro-micro decomposition (see \cite{Sluse2007, Hutsemekers2010, Sluse2012b}), consists in linearly combining the spectra with two multiplicative factors: a factor $\mu$, which is the amplitude of microlensing in the continuum, and a factor $M$ which accounts for the macro-magnification associated to the strong-lensing effect, and which affects identically all the emitting regions. By linearly combining the spectra, it is possible to isolate the fraction of the flux $F_M$ coming from regions too large to be microlensed, from the fraction of the flux $F_{M\mu}$ microlensed at most like the continuum. This method is well suited to unveil microlensing of the broad emission lines, independently of any model of the emission. We show in Fig.~\ref{fig:MmD}, an application of the MmD on HE0047-1756. Other examples and tests of this method can be found in \cite{Sluse2012b}. 


One important information on the AGN structure we want to derive is the ``size'' of the emitting regions. This information is retrieved with generally large uncertainties from spectra of the lensed images obtained at a single epoch (i.e. smaller is the amplitude of microlensing, larger will be the uncertainties). However, this procedure applied to several systems enables one to derive average sizes for samples of objects. In order to derive sizes of a single object with smaller uncertainties, it is necessary to obtain monitoring data which provide lightcurves for each lensed image. A possible strategy to derive sizes from these lightcurves has been described in \cite{Kochanek2004}. (i) First, one calculates differential lightcurves{\footnote{Instead, one can also use a model for the quasar intrinsic variability and work on individual images.}} (corrected for a time delay if any) between pairs of images in order to remove intrinsic variability from the signal. (ii) Second, micro-amplification maps associated to the macro-images are generated. These maps correspond to the patterns of caustics produced by the stars in the lensing galaxy projected onto the source. By convolving the light distribution of the source with the pattern, and moving the source through these maps, one simulates microlensing lightcurves. Note that these maps depend on the fraction of compact objects $f_*$ in the lensing galaxy, at the position of the lensed images. The latter is chosen depending of our knowledge of the lens. Alternatively, maps with different $f_*$ are used, and results are marginalized over the different values of $f_*$. (iii) The third step is to find a representative sample of lightcurves which ``fit'' the data. Those lightcurves are characterized by several parameters, the most relevant ones being the size of the source and the relative transverse velocity of the microlenses. From step (iii) it is possible to derive the posterior probability distribution associated to the parameters of interest, like the size of the emitting region. 



\section{AGNs' size from microlensing: the test-case of the Einstein cross} 
\label{sec:EC}

The Einstein-Cross $\equiv$ Q2237+0305 is the archetype microlensed quasar. The source ($z_s=1.695$) is located behind the bulge of a nearby spiral galaxy located at $z_l=0.0395$. This particular configuration leads to a high relative transverse velocity which translates into ''quick'' microlensing events lasting typically several months. Because of these characteristics, there have been many microlensing studies of this system. Based on photometric monitoring from the X-ray to the optical wavelengths, sizes of the corresponding emitting regions have been retrieved (e.g. \cite{Poindexter2010, Mosquera2013}). Photometric monitoring allows one to derive the size of the continuum emission. In order to study the emission arising from the BLR, spectrophotometry is needed. 

\begin{figure}[tb]
\begin{center}
\includegraphics[width=9cm]{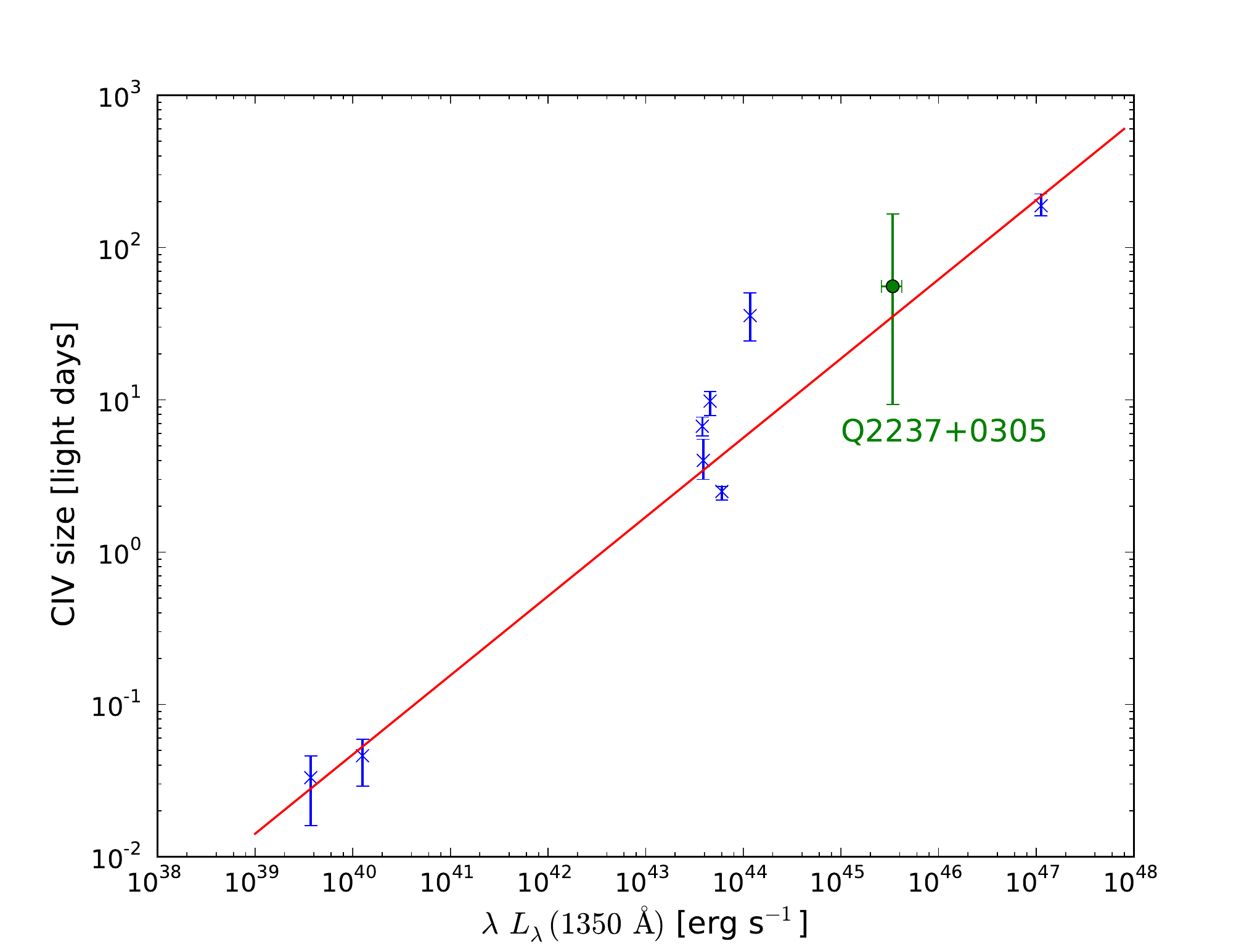} 
\caption{ $R_{CIV}$-L diagram. Reverberation mapping measurements with 'x' symbol ~\cite{Peterson2005, Kaspi2007}, together with the microlensing-based size for Q2237+0305 \cite{Sluse2011}. A fit to the reverberation mapping data from \cite{Kaspi2007} is depicted with a solid line (slope $\alpha = 0.52$).}
\label{fig:revsize}
\end{center}
\end{figure}

\cite{Eigenbrod2008a} performed the first spectroscopic monitoring of Q2237+0305 using the MXU mode of the FORS instrument at the VLT. At a cadence of about one observation per month, they obtained spectra of the four lensed images, and of several field stars for about 3 years. They took advantage of the field stars to deconvolve the spectra of the lensed images (blended with the lensing galaxy), and perform a simultaneous flux calibration. The analysis of the data revealed a strong microlensing variability of the lensed images A and B, and much smaller microlensing variations in images C and D. Two striking features were unveiled in those spectra: first, the spectral slope became steeper during the microlensing amplification events; second, the emission lines in image A were microlensed during the whole observing period \cite{Eigenbrod2008a}. The chromatic trend observed in the continuum during a microlensing event reflects the increasing size of the continuum emitting region from blue to red. Such a color change had been reported previously (e.g. \cite{Anguita2008}). The reason is that the accretion disc emission is more compact at bluer wavelengths than at redder ones. This effect has been used to constrain the temperature profile of the disc. For this purpose, each individual spectrum has then been modeled as a sum of a power law continuum and of broad emission lines (including an empirical template for the \FeII\,blend). This procedure enabled us to separate the flux of the continuum and of the broad lines. A lightcurve for the continuum emission in six different wavelength ranges and in the BELs has been derived. The technique outlined in Sect.~\ref{sec:method} has then been used to model these lightcurves, in combination with higher cadence lightcurves from a dense photometric monitoring. This procedure led to a constraint on the size of the continuum emission, $R$, as a function of wavelength: $R \propto \lambda^\eta$. A value $\eta = 1.2 \pm 0.3$ has been obtained. This index is related to the temperature profile of the accretion disc and compatible with the predictions of a Shakura-Sunyaev accretion disc model ($\eta = 1.3$). A similar procedure applied to the lightcurves of images A and D led to the measurement of the size of the \civ\,emitting region. We derived a most likely size (half-light radius) for the \civ\,emitting region (assuming an average mass of the microlenses $\left< M \right> \sim 0.3 M_{\odot}$) of $R_{CIV} \sim 66^{+110}_{-46}$ light-days = 0.06$^{+0.09}_{-0.04}$ pc = 1.7$^{+2.8}_{-1.1}$\,10$^{17}$ cm (at 68.3\% CI). Similar values were obtained for the \ciii\,emitting region~\cite{Sluse2011}. We plot the size we derived for the \civ\,emitting\,region on a Radius-Luminosity diagram obtained from reverberation mapping in low redshift systems \cite{Peterson2005, Kaspi2007}. Our estimate appears in very good agreement with the relation observed at low redshift (Fig.~\ref{fig:revsize}). 

The analysis of the spectra using the MmD decomposition reveals that only a fraction of the broad line is affected by microlensing. The modeling of the emission lines with a sum of gaussian components of different widths confirms this finding. In particular, the \civ\,line is well reproduced with a gaussian G1 with FWHM$\sim$6300\,km/s (of slightly varying width over the monitoring period) and of a narrower component G2 with FWHM$\sim$2600\,km/s. The broadest component G1 is found to produce the main contribution to the microlensing signal. The microlensing size derived for that region is about four times smaller than the one derived from the whole line. Another interesting output of the procedure is the ability to derive the size ratio between the broad line region and the continuum emission. For the \civ\, emission, we found that the broad component of the line arises from a region only four time larger than the region emitting the $V$-band continuum, while the narrower component arises from a region 25 times larger than the continuum emitting region.  

\section{Microlensing in other systems}
\label{sec:BLR}

Nowadays, microlensing of the continuum emission has been identified in almost all the systems for which good quality spectra of the individual lensed imaged have been obtained. This observational result is expected because the continuum size is typically less than one tenth of an Einstein radius, and the optical depth of microlensing is close to one. The larger broad line region is expected to be affected to a smaller extent by microlensing, and has commonly received less attention. Recently, we performed a first systematic search for microlensing of the BLR in a sample of archive spectra of 13 lensed systems{\footnote{The spectra are available via CDS \url{http://vizier.cfa.harvard.edu/viz-bin/VizieR?-source=J/A+A/544/A62} and \url{http://dc.g-vo.org/mlqso/q/web/form}}} \cite{Sluse2012b}. The sample was composed of objects in the bolometric luminosity range $10^{44.7}-10^{47.4}$\,erg/s and black hole masses $10^{7.6}-10^{9.8}\,M_{\odot}$. Those systems were targeted originally to measure the redshift of the lensing galaxy, but not for microlensing studies. Although it might not be free of selection effects, this sample should be representative of the frequency and characteristics of microlensing in lensed quasars.
Using the MmD decomposition technique (Sect.~\ref{sec:method}), we unveiled microlensing-induced deformation of the broad emission lines independently of any spectral modeling. An example of decomposed spectrum is shown in Fig.~\ref{fig:MmD} (see \cite{Sluse2012b} for other systems). The main broad lines visible in the spectra were \MgII\,and \CIII. We found these lines to be microlensed in about 70-80\% of the systems. The signal was found to be in general of low amplitude, leading to typically 10\% of line-flux variation. Very interestingly, we found that, in general, one wing of the emission line (either the blue or the red wing) was much more affected by microlensing than the other one. This finding contrasts with most of the previous detections of microlensing of emission lines, found to affect both wings with similar amplitude. This simple observation is another proof that the BLR does not have in general a spherically symmetric geometry and velocity field. Further insights on this question may come from simulated microlensing variations assuming different geometries for the BLR. 

The microlensing technique is a powerful tool to probe the AGN structure. It nicely complements existing methods, and is currently the only one which can be applied to objects of intrinsically large luminosity (i.e. $L_{bol} > 10^{45}$\,erg/s), and with relatively large redshifts ( i.e. 0.5 $< z < $ 4.5, the typical range of known microlensed AGNs). 


\end{document}